\documentclass[twocolumn,prl,showpacs,amsmath,amssymb]{revtex4}

\usepackage{graphicx}
\usepackage{dcolumn}
\usepackage{bm}

\begin{document}

\title{$Q^2$ Evolution of the Neutron Spin Structure Moments using a $^3$He Target.}

\author{
M.~Amarian,$^{\nyerevan}$
L.~Auerbach,$^{\ntemple}$
T.~Averett,$^{\njlab,\nwm}$
J.~Berthot,$^{\nclermont}$
P.~Bertin,$^{\nclermont}$
B.~Bertozzi,$^{\nmit}$
T.~Black,$^{\nmit}$
E.~Brash,$^{\nregina}$
D.~Brown,$^{\nmaryland}$
E.~Burtin,$^{\nsaclay}$
% ...
}
\affiliation{
\baselineskip 2 pt
\vskip 0.3 cm
{\rm (Jefferson Lab E94010 Collaboration)} \break
\vskip 0.1 cm
\centerline{{$^{\ncaltech}$California Institute of Technolgy, Pasadena, California 91125}}
\centerline{{$^{\nchungham}$Chungnam National University, Taejon 305-764, Korea}}
\centerline{{$^{\nhampton}$Hampton University, Hampton, Virginia 23668}}
\centerline{{$^{\nclermont}$LPC IN2P3/CNRS, Universit\'e Blaise Pascal, F--63170 Aubi\`ere Cedex,
France}}
\centerline{{$^{\ninfn}$Istituto Nazionale di Fiscica Nucleare, Sezione Sanit\`a, 00161 Roma, Italy}}
\centerline{{$^{\njlab}$Thomas Jefferson National Accelerator Facility, Newport News, Virginia
23606}} 
\centerline{{$^{\nkent}$Kent State University, Kent, Ohio 44242}}
\centerline{{$^{\nkentucky}$University of Kentucky, Lexington, Kentucky 40506}}
\centerline{{$^{\nkharkov}$Kharkov Institute of Physics and Technology, Kharkov 310108, Ukraine}}
\centerline{{$^{\nmaryland}$University of Maryland, College Park, Maryland 20742}}
\centerline{{$^{\nmit}$Massachusetts Institute of Technology, Cambridge, Massachusetts 02139}}
\centerline{{$^{\numass}$University of Massachusetts-Amherst, Amherst, Massachusetts 01003}}
\centerline{{$^{\nunh}$University of New Hamphsire, Durham, New Hamphsire 03824}}
\centerline{{$^{\nodu}$Old Dominion University,  Norfolk, Virginia 23529}}
\centerline{{$^{\nprinceton}$Princeton University, Princeton, New Jersey 08544}}
\centerline{{$^{\nregina}$University of Regina, Regina, SK S4S 0A2, Canada}}
\centerline{{$^{\nrutgers}$Rutgers, The State University of New Jersey, Piscataway, New Jersey
08855}}
\centerline{{$^{\nsaclay}$CEA Saclay, DAPNIA/SPHN, F--91191 Gif/Yvette, France}}
\centerline{{$^{\nsyracuse}$Syracuse University, Syracuse, New York 13244}}
\centerline{{$^{\ntemple}$Temple University, Philadelphia, Pennsylvania 19122}}
\centerline{{$^{\ntohoku}$Tohoku University, Sendai 980, Japan}}
\centerline{{$^{\nuva}$University of Virginia, Charlottesville, Virginia 22904}}
\centerline{{$^{\nwm}$The College of William and Mary, Williamsburg, Virginia 23187}}
\centerline{{$^{\nyerevan}$Yerevan Physics Institute, Yerevan 375036, Armenia}}
}

\newcommand{\ncaltech}{1}
\newcommand{\nchungham}{2}
\newcommand{\nhampton}{3}
\newcommand{\nclermont}{4}
\newcommand{\ninfn}{5}
\newcommand{\njlab}{6}
\newcommand{\nkent}{7}
\newcommand{\nkentucky}{8}
\newcommand{\nkharkov}{9}
\newcommand{\nmaryland}{10}
\newcommand{\nmit}{11}
\newcommand{\numass}{12}
\newcommand{\nunh}{13}
\newcommand{\nodu}{14}
\newcommand{\nprinceton}{15}
\newcommand{\nregina}{16}
\newcommand{\nrutgers}{17}
\newcommand{\nsaclay}{18}
\newcommand{\nsyracuse}{19}
\newcommand{\ntemple}{20}
\newcommand{\ntohoku}{21}
\newcommand{\nuva}{22}
\newcommand{\nwm}{23}
\newcommand{\nyerevan}{24}

\date{\today}

\begin{abstract}
We have measured the spin structure functions $g_1$ and $g_2$ of
$^3$He in a double-spin experiment by inclusively scattering polarized  electrons
at energies ranging from 0.862 to 5.058~GeV off a polarized $^3$He target  at a
15.5$^{\circ}$ scattering angle. Excitation energies covered the resonance and the onset 
of the deep inelastic regions.  We have determined for the first time the 
$Q^2$ evolution of $\Gamma_1(Q^2)=\int_0^{1} g_1(x,Q^2) dx$,  $\Gamma_2(Q^2)=\int_0^1
g_2(x,Q^2) dx$  and $d_2 (Q^2) = \int_0^1 x^2[ 2g_1(x,Q^2) + 3g_2(x,Q^2)] dx$ for the neutron in the range 
0.1~GeV$^2$ $\leq Q^2 \leq $ 0.9~GeV$^2$ with good precision. 
$ \Gamma_1(Q^2)$ displays a smooth variation from high to low
$Q^2$. The Burkhardt-Cottingham sum rule holds within uncertainties and 
$d_2$ is non-zero over the measured range. 
\end{abstract}


\maketitle

During the past twenty five years, our understanding of Quantum Chromodynamics
(QCD)  has advanced through the study of the spin structure of the nucleon. 
Measurements of the nucleon spin structure functions $g_1$ and $g_2$ in deep
inelastic  lepton scattering (DIS) were used to unravel the spin structure of the
nucleon  in terms of its constituents, quarks and gluons, and test QCD.  Among the
important  results are  the finding that only a small fraction (about $ 20 \%$) of
the nucleon spin is accounted for by the spin of quarks~\cite{Hug:99}, and 
the test of the Bjorken sum rule~\cite{Bjo:66}, a fundamental sum rule of QCD,  to
better than 10\%.  To make this latter test possible and determine the quark
contribution to the total spin,  it was essential to calculate the corrections 
necessary to evolve the Bjorken sum rule and the first moment of
$g_1$~\cite{Eja:74} to $Q^2$ values accessible experimentally. To this end the sum
rule, originally derived in the limit $Q^2\rightarrow \infty $ using current algebra,  was
generalized using the technique of operator product expansion (OPE) in
QCD~\cite{Kod:79,Kod:80,Lar:91,Shu:82}. In this connection, it was also realized 
that the Bjorken sum rule was  a special case of a more general sum rule known as the extended
Gerasimov-Drell-Hearn (GDH) sum rule~\cite{Ji:99} which spans the full range of momentum transfer
from $Q^2$ = 0 to $Q^2\rightarrow \infty$.

At small $Q^2$, after subtracting the elastic 
contribution, $\bar \Gamma_1(Q^2)=\Gamma_1(Q^2) -  \Gamma_1(Q^2)^{\rm elastic}$, is
linked to the anomalous magnetic moment of the nucleon $\kappa$ by  
\begin{equation}
\bar \Gamma_1(Q^2) =\int_0^{x_0} g_1(x,Q^2) dx =  -{Q^2\over 8M^2}\kappa^2 +  O
\biggl ({Q^4\over M^4}\biggl ), 
\label{gdh:1}
\end{equation} where $x_0$ coincides with the nucleon pion threshold. The first
term in the right hand side of (\ref{gdh:1}) corresponds to the original GDH sum
rule prediction~\cite{Ger:65}. The next term has been  evaluated by Ji {\it et al.} using a
heavy-baryon chiral perturbation  theory (HB$\chi$PT)~\cite{Ji:99,Ji:00} and by Bernard
{\it et al.} using a covariant chiral perturbation theory 
($\chi$PT)~\cite{Ber:02,Ber:03}.

At large $Q^2$ ($Q^2\gg \Lambda_{QCD}^2$) $\Gamma_1(Q^2)$ is expressed 
in terms of a twist expansion~\cite{Ji:93,Ji:94}:
\begin{equation}
\Gamma_1(Q^2) = \frac{1}{2}a_0 + \frac{M^2}{9Q^2}\biggl (a_2 + 4d_2 +4f_2 \biggr) 
+ O\biggl (\frac{M^4}{Q^4}\biggr ),
\end{equation}
where $a_0$ is the dominant, leading twist contribution. It is determined, apart
from QCD radiative corrections~\cite{Lar:91},  by the triplet $g_A$ and octet $a_8$
axial charges and the net quark spin contribution to the total nucleon spin. These
axial charges are extracted from measurements of  the neutron and hyperons weak
decay measurements~\cite{Clo:94}. Here 
$a_2$ is a second moment of the $g_1$ structure function and arises from the target
mass correction~\cite{Ji:94}. The quantities $d_2$ and  $f_2$ are the twist-3 
and the twist-4 reduced matrix elements. These matrix elements contain
non-trivial quark-gluon interactions beyond the parton model. A first attempt at
extracting $f_2$  has been carried out by Ji and Melnitchouk~\cite{Ji:97} using the
world data but with poor statistics below $Q^2$ = 1 GeV$^2$. In QCD, $d_2$ and
$f_2$ can be expressed as linear combinations of the  induced color electric and
magnetic polarizabilities $\chi_E$ and
$\chi_B$~\cite{Ste:95,Ji:95} when a nucleon is polarized. The above twist expansion
may be valid down to $Q^2 = 0.5~$GeV$^2$ if higher order terms are small.

We define $d_2$ as the second moment of a particular 
combination of the measured $g_1$ and $g_2$ structure functions:
\begin{eqnarray}
d_2(Q^2) &=& \int_0^{1} x^2 \bigl [2g_1(x,Q^2)+3 g_2(x,Q^2) \bigr] dx \\
&=& {3}\int_0^{1} x^2 \biggl [g_2(x,Q^2)-g_2^{WW}(x,Q^2)\biggr ]dx \nonumber
\end{eqnarray}
where $g_2^{WW}$, known as the Wandzura-Wilczek~\cite{Wan:77} term, 
depends only on $g_1$
\begin{equation}
 g_2^{WW}(x,Q^2) = -g_1(x,Q^2) + {\displaystyle \int_x^1 \frac{g_1(y,Q^2)}{y} dy}. 
\end{equation}
The quantity $d_2$ reduces to a twist-3 matrix element at large $Q^2$ where
an OPE expansion becomes valid.

The advantages of measuring higher moments of the spin structure functions are 
twofold; 1) the  kinematical region experimentally accessible gives most 
of the contribution to these moments, and 2) the matrix elements in the OPE
of these moments can be calculated using lattice QCD~\cite{Goc:01}. 

Most of the previous measurements of $\Gamma_1$, $\Gamma_2$ and $d_2$  were performed
at $Q^2$  well above 1~GeV$^2$ where the higher-twist contributions are small compared
to the precision of the experiments. However,  a good precision test of
OPE requires precision data of $\Gamma_1^n$ starting from 
$Q^2$ of about 0.5~GeV$^2$  where multiparton interactions are important.
From $Q^2 = 0$ to perhaps $Q^2$ = 0.2~GeV$^2$, the moments predicted 
by the sum rules (e.g. spin polarizabilities etc...) can be calculated in
chiral perturbation  theory ($\chi$PT)~\cite{Ji:00,Ber:03} and can be tested
against experiments. We do not expect OPE or $\chi$PT to be  valid in the complete
range of $Q^2$; however, with time,  lattice QCD may bridge the gap
between these two limits.  

Finally, the $g_2$ structure function itself is predicted to obey 
the Burkhardt-Cottingham (BC) sum rule
\begin{equation}   
\Gamma_2 (Q^2) = \int_0^1 g_2(x,Q^2)~dx=0\ 
\label{eq:bc}
\end{equation}
which was derived from the dispersion relation and the  asymptotic behavior of the
corresponding Compton amplitude~\cite{Bur:70}.  This sum rule is also expected to
be valid at all $Q^2$ and does not follow from the OPE. 
It is a super-convergence relation based on Regge asymptotics as discussed in the review paper 
by Jaffe~\cite{Jaf:90}. Many scenarios which could invalidate this sum
rule have been  discussed in the literature~\cite{Iva:99,Ans:95,Jaf:91}. However,
this sum rule was confirmed in perturbative QCD at order 
$\alpha_s$ with a $g_2 (x,Q^2)$ structure function for  a quark target
\cite{Alt:94}. Surprisingly  the first precision measurement of $g_2$ at SLAC~\cite{Ant:02}  
at $Q^2$ = 5~GeV$^2$ but within a limited range of
$x$ has revealed a violation of this sum rule on the proton at the level of three
standard deviations. In contrast,  the neutron sum rule is poorly measured but
consistent with zero at the one  standard deviation. 

In this paper, we present measurements of the spin structure functions
$g_1$ and $g_2$  of $^3$He and the determination of 
$\Gamma_1^n(Q^2)$, $\Gamma_2^n(Q^2)$ and $d_2^n(Q^2)$ for the neutron 
below $Q^2$ = 1~GeV$^2$. The data on $\Gamma_1^n(Q^2)$ and $d_2^n(Q^2)$ provide a test
of the latest results in $\chi$PT and permit a better extraction of $f_2$.  
The data on $\Gamma_2^n(Q^2)$ allows us 
to make a more precise test of the neutron BC sum rule at  $Q^2 < 1$ GeV$^2$.  

The features of JLab experiment E94-010, from which these results are derived,
 were discussed  in a previous determination of
$\sigma_{TT'}$, the transverse-transverse virtual  photoabsorption cross
section~\cite{Ama:02}. We measured the inclusive scattering of longitudinally 
polarized electrons from a polarized $^3$He target in Hall A at the Thomas 
Jefferson National Accelerator Facility (JLab).  Data were collected at six incident 
beam energies: 5.058, 4.239, 3.382,
2.581, 1.718, and 0.862~GeV, all at a nominal scattering angle of 15.5$^{\circ}$. 
The  measurements covered values of the invariant mass $W$ from the quasielastic
peak (not discussed in this paper), through the resonance region and continuum.  
Data were taken for  both longitudinal and transverse target polarization
orientations.  Both spin asymmetries and absolute cross sections were measured. 
More details can be found in~\cite{Site:94}.

In the Born approximation $g_1(x,Q^2)$ and $g_2(x,Q^2)$ are evaluated
by  combining data taken with opposite electron beam helicity and 
parallel or perpendicular target spin with respect to the beam direction. 

\begin{eqnarray}
 g_1 &=& {MQ^2\nu \over 4\alpha_e^2}{ E\over E'}{1\over E+E'} \biggl [\Delta\sigma_\|+
\tan(\theta/2)\Delta\sigma_\bot\biggr ] \\
 g_2 &=& {MQ^2\nu^2 \over 4\alpha_e^2}{1\over 2E'(E+E')} \biggl [-\Delta\sigma_\| +
\frac{E+E'\cos{\theta}}{E'\sin{\theta}}\Delta\sigma_\bot\biggr ], \nonumber
\label{eq:g1g2}
\end{eqnarray}
where $\Delta\sigma_{\parallel (\perp)} = {{d^2\sigma^{\downarrow\Uparrow
(\Rightarrow)}}/{d\Omega
dE^{\prime}}}-{{d^2\sigma^{\uparrow\Uparrow(\Rightarrow)}}/{d\Omega dE^{\prime}}}$
is the difference of cross sections  for the case in which the target spin  is 
aligned parallel (left-perpendicular) to the beam momentum. Here $\alpha_e$ is the
electromagnetic coupling constant, $\theta$ is the scattering angle, $M$ is the
nucleon mass,
$\nu$ is the transferred energy and $E$ and $E^{\prime}$ are the initial and final
energies of the incident and scattered electron respectively.

\begin{figure}[ht!]
\begin{center}
\centerline{\includegraphics[scale=0.60]{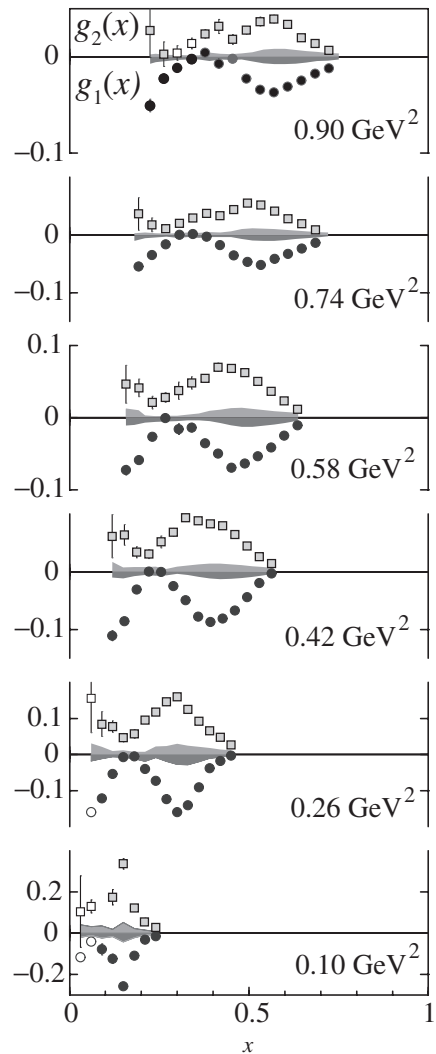}}
\end{center}
\caption{$g_1$ (dark filled circles) and $g_2$ (grey filled squares) of $^3$He are
plotted as a function of the Bjorken variable $x$ for six values of
$Q^2$. The points shown with filled (open) circles or  squares were
determined by interpolation (extrapolation).}
\label{fig:gone_gtwo_Q}
\end{figure}

The results of $g_1$ (circles) and $g_2$ (squares) for $^3$He are shown
in Fig.~\ref{fig:gone_gtwo_Q} as a function of $x$ for six 
values of $Q^2$ in the range  $0.1~{\rm GeV}^2 \le Q^2 \le 0.9~{\rm
GeV}^2$. These structure functions were  evaluated at constant $Q^2$ from those
measured at fixed incident beam energies and angle by interpolation (filled symbols), and for a few
points (open symbols), extrapolation.  The error bars represent the uncertainty due to statistics
only, and the grey bands indicate the uncertainty due to systematic errors. The
systematic errors result from relative uncertainties of about  5\%  in the absolute
cross sections, 4\% in the target polarization, 4\% in the beam polarization, and
20\% in the radiative corrections at every but the lowest incident beam energy where it is 
40\%. They also include a contribution from interpolation and extrapolation. 

We notice a peak in $g_1$ and $g_2$ due to the $\Delta_{1232}$
resonance,  which decreases in magnitude with increasing $Q^2$. 
In the vicinity of that resonance $g_1$ and $g_2$ are of almost equal
amplitude and opposite sign. This is consistent with the fact that
the $\Delta(1232)$ is an M1 resonance, leading us to expect
$\sigma_{LT'} \propto (g_1 + g_2)$\cite{Dre:01} to be highly supressed. 
We note also that in a region dominated by the coherent behavior of quarks 
and gluons (constituent
quarks instead of current quarks) the Wandzura-Wilczek relation derived in DIS still
holds, perhaps pointing to  the role of quark-hadron duality in
$g_1$.

The integral $\bar \Gamma_1(Q^2)$ for $^3$He was computed for each value of $Q^2$  using
limits of integration extending from the nucleon pion threshold to a value of $x$
corresponding to $W=2.0\,{\rm GeV}$. To extract $\bar \Gamma_1^n(Q^2)$ we followed the
prescription suggested by Ciofi degli Atti and Scopetta in~\cite{Cio:97}, where it
is found, within the Impulse Approximation, that nuclear effects are quite
significant when extracting the spin structure functions, but they reduce to at most 10~\%
 in the extraction of  $\bar \Gamma_1^n(Q^2)$.

\begin{figure}[ht!]
\begin{center}
\centerline{\includegraphics[scale=0.50]{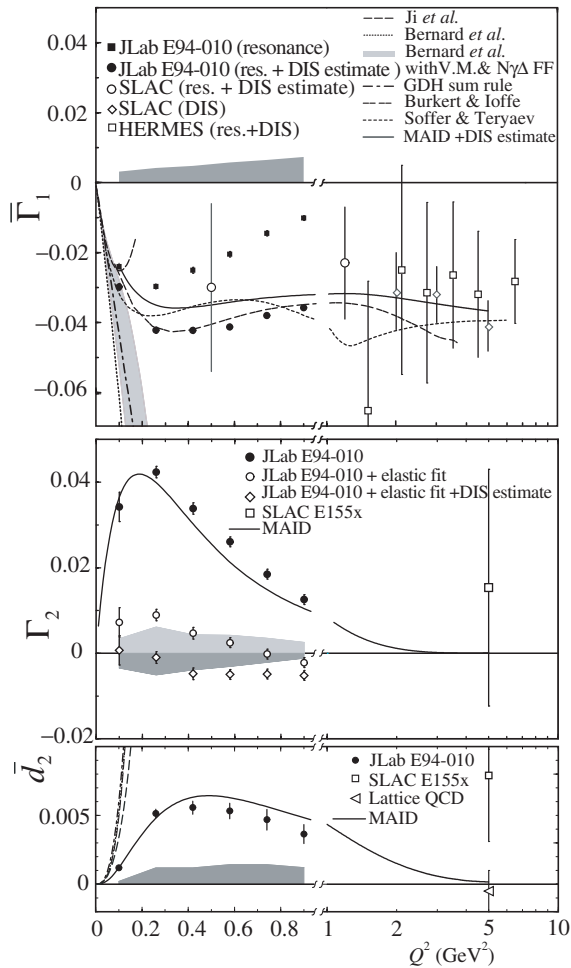}}
\end{center}
\caption{Results of  $\overline{\Gamma}_1^n$ (top panel),  $\Gamma_2^n$ (middle
panel), $\bar d_2^n$ (bottom panels )  along with the world  data from DIS and 
theoretical predictions (see text).} 
\label{fig:gdh_int}
\end{figure}

The measured values of  $\bar \Gamma_1^n(Q^2)$ (solid squares) are shown in Fig.~\ref{fig:gdh_int}  (top
panel) at six values of $Q^2$ along with the world data (open symbols) from SLAC~\cite{SLAC:all} and 
Hermes~\cite{Her:98}. To obtain the total result of $\bar \Gamma_1^n(Q^2)$ (solid circles )  
an estimated strength in the DIS range 4 GeV$^2 < W^2 < 1000$ GeV$^2$ using the 
parametrization~\cite{Tho:00,Der:01} 
was added to the data obtained in the measured region (solid squares).  
The size of each symbol indicates the statistical
uncertainty  while the systematic uncertainties are illustrated with the dark grey band
along the horizontal axis and include the uncertainty of the DIS contribution.  
We point out that at $Q^2 > 1$ GeV$^2$ the elastic contribution is negligible and therefore 
$\bar \Gamma(Q^2) = \Gamma(Q^2)$. 

At low $Q^2$ we show several $\chi$PT calculations; by Bernard {\it et al.}~\cite{Ber:02} 
without vector mesons (dotted line), by Bernard {\it et al.}~\cite{Ber:03} including vector mesons 
and $\Delta$ degrees of freedom (grey band) and by Ji {\it et al.}~\cite{Ji:99} 
using the HB$\chi$PT (dashed line). In the calculation of Bernard {\it et al.}~\cite{Ber:03}
the grey band  shows a range of results due to the uncertainty in the N$\gamma \Delta$ form factor. 
 The latter calculation overlaps with a data point at $Q^2$ = 0.1 GeV$^2$.
The GDH prediction is the slope depicted by the dot-dashed line at low $Q^2$.

At moderate and large $Q^2$  we show the MAID calculation~\cite{Dre:01} (used to
evaluate the resonance contribution) combined with the DIS estimate from  Bianchi and Thomas~\cite{Tho:00} 
(solid line). The other calculations shown are by Soffer and Terayev~\cite{Sof:02}
(short-dashed line) and  by Burkert and Ioffe~\cite{Bur:92} (double-dashed line).

While the MAID and  Soffer and Terayev calculations are disfavoured, the result from the Burkert and
Ioffe calculation agrees well with the data.  More importantly, our data above $Q^2$ = 0.5 GeV$^2$ combined 
with the world data and future planned measurements~\cite{Mez:03,JLa:00} in the range
1 GeV$^2 < Q^2 < 5$ GeV$^2$  will permit, in the
framework of QCD and with better experimental constraints, to repeat the 
extraction of $f_2 (Q^2)$ higher-twist matrix
element performed by Ji and Melnitchouk. 

We plot in Fig.~\ref{fig:gdh_int} (middle panel)  $\Gamma_2^n$ in the measured 
region (solid circles) and after adding the elastic contribution (open circles)
evaluated using the Mergell {\it et al.}~\cite{Mer:96} parameterization of $G_M^n$ and 
$G_E^n$. The open diamonds correspond to the results obtained after adding 
to the open circles an estimated DIS contribution assuming $g_2 = g_2^{WW}$ using
the same method as described in~\cite{Ant:02}. Nuclear corrections were performed  
following a procedure similar to that
used in extracting $\Gamma_1^n$ including $Q^2$ dependent effects~\cite{Mel:03}. 
The solid line is the resonance contribution evaluated using MAID.
The positive light grey band corresponds to the total experimental systematic
errors. The negative dark band is our best estimate of the systematic error 
for the low $x$ extrapolation. 
 
Our neutron results (open diamonds) are consistent with the BC sum rule to 
within 1.7 standard deviations over the measured $Q^2$ range. 
The SLAC E155x collaboration~\cite{Ant:02} previously reported 
a neutron result at high $Q^2$ (open square), where the elastic contribution 
is negligible,  consistent with zero but with a rather 
large error bar. On the other hand, they reported a proton result which deviates 
from the BC sum rule by 2.8 standard deviations. Their quoted error bar in this case is
3 times smaller than that of the neutron result but still large. 

In Fig.~\ref{fig:gdh_int} (bottom panel), $\bar d_2(Q^2) = d_2(Q^2) - d_2^{\rm {elastic}}(Q^2)$  
is shown at several values of $Q^2$. The results of this experiment are the solid circles. 
The grey band represents their corresponding systematic uncertainty. The SLAC  
E155x~\cite{Ant:02} neutron result (open square) is also shown. The solid line is the
MAID calculation containing only the resonance contribution. 
At low $Q^2$ the HB$\chi$PT calculation\protect\cite{Van:02} (dashed line) and the covariant 
$\chi$PT (dotted line) are shown. The two latter calculations overlap in the $Q^2$ region shown. 
Furthermore, calculations of the covariant $\chi$PT with the vector mesons contribution (dot-dashed line) 
and the $\Delta$ degrees of freedom (long dashed line) are  reported~\cite{Ber:03} but are 
too close to the former  $\chi$PT curves to be clearly seen at this scale.

The lattice prediction~\cite{Goc:01} at $Q^2$ = 5~GeV$^2$ for the neutron $d_2$ matrix
element is negative but close to zero. We note that all models (not shown at
this scale) predict a negative or zero value at large $Q^2$. At moderate $Q^2$ 
our data show a positive $\bar d_2^n$ and indicate  a slow decrease with $Q^2$.

In conclusion, we have made the first measurement of $\Gamma_1^n(Q^2)$, 
$\Gamma_2^n(Q^2)$ and $d_2^n(Q^2)$ of the neutron from a $Q^2$ regime where a
twist expansion analysis is appropriate to a regime where $\chi$PT can be
tested.  The BC sum rule for the neutron is verified within errors in the
intermediate range of $Q^2$  due to a cancellation between the resonance and the
elastic contributions.  Our $\bar d_2^n$ results 
suggest a positive contribution of the resonance region up to $Q^2$ = 1 GeV$^2$
of a size comparable to the SLAC E155x result.  With time we
expect our data, combined with future measurements to provide a  challenging
test of increasingly precise lattice QCD predictions.

This work was supported by the U.S. Department of Energy (DOE), the U.S.
National  Science Foundation, the European INTAS Foundation, and the French
CEA, CNRS, and  Conseil R\'egional d'Auvergne. The Southeastern Universities
Research Association  (SURA) operates the Thomas Jefferson National Accelerator
Facility for the DOE under  contract DE-AC05-84ER40150.

\end{document}